\documentclass[aps,prx,amsmath,nofootinbib,longbibliography,amssymb,superscriptaddress,twocolumn,10pt]{revtex4-2}

\usepackage[english]{babel}
\usepackage{graphicx}
\usepackage[normalem]{ulem}
\usepackage{soul}
\usepackage{float}
\usepackage[svgnames]{xcolor}
\usepackage{comment}
\usepackage[colorlinks, linkcolor=NavyBlue , citecolor= NavyBlue,urlcolor = NavyBlue, breaklinks=true]{hyperref}
\usepackage{verbatim}
\usepackage{mathtools}
\usepackage{esint}
\usepackage{marginnote}

\newcommand{\RNum}[1]{\uppercase\expandafter{\romannumeral #1\relax}}

\begin{document}

\title{Collective Modes of Fractional States Beyond Landau Levels}
\author{Peleg Emanuel}
\affiliation{Department of Condensed Matter Physics, Weizmann Institute of Science, Rehovot 76100, Israel}

\author{Yuval Oreg}
\affiliation{Department of Condensed Matter Physics, Weizmann Institute of Science, Rehovot 76100, Israel}

\date{\today}

\begin{abstract}
Collective modes provide a powerful probe into the properties and stability of topologically ordered and strongly interacting phases of matter, such as fractional quantum Hall states. Yet, the description of fractional-state excitations away from the Landau-level limit remains a formidable challenge. We develop a framework for describing collective excitations in the strong-coupling regime of coupled-wire constructions and show that they emerge as solitons and breathers of effective sine-Gordon theories. This approach provides a systematic route to studying collective modes in a broad class of fractional states.
Applied to a coupled-wire construction of the fractional quantum Hall effect, the framework captures key features of the magnetoroton branch. It naturally extends to fractional Chern insulators, demonstrating that magnetoroton-like collective modes persist well beyond the Landau-level limit. We find that the magnetoroton gap increases as the Fubini–Study metric approaches the ideal (trace-condition) limit, highlighting the important role of quantum geometry in the stability of fractionalized phases and their collective excitations. The same trend favors the stability of the long-wavelength quadrupolar mode associated with the chiral graviton. Together, these results identify a distinct interplay between fractionalization and quantum geometry, and provide an analytical framework for studies of collective excitations in fractional topological matter.
\end{abstract}
\maketitle

\section{Introduction}

Fractional Chern insulators (FCIs), lattice analogs of the fractional quantum Hall effect (FQHE), have attracted considerable attention in recent years \cite{liuRecentDevelopmentsFractional2024}.
FCIs generalize the FQHE, and, as suggested theoretically over a decade ago~\cite{kapitExactParentHamiltonian2010, qiGenericWaveFunctionDescription2011, regnaultFractionalChernInsulator2011, shengFractionalQuantumHall2011, tangHighTemperatureFractionalQuantum2011, neupertFractionalQuantumHall2011}
and recently verified in experiments~\cite {zengThermodynamicEvidenceFractional2023, caiSignaturesFractionalQuantum2023, parkObservationFractionallyQuantized2023, luFractionalQuantumAnomalous2024,
luExtendedQuantumAnomalous2025,butler13Fractional2026},
may be realized in the absence of a magnetic field. Thus, they may enable practical innovations obstructed by high magnetic fields. 
While FCIs exhibit the hallmark feature of the FQHE, a quantized fractional Hall conductance accompanied by a vanishing longitudinal resistance, it is still unclear how much of the extensively studied phenomenology of the FQHE carries over to their lattice analogs.
For instance, the lowest-lying neutral collective mode of the FQHE, commonly known as the magnetoroton branch~\cite{girvinCollectiveExcitationGapFractional1985, girvinMagnetorotonTheoryCollective1986}, has been studied in depth. Does a similar excitation branch appear for FCIs?

The answer to this question has several implications. First, the FQH magnetoroton achieves its energy minimum at a finite momentum, and hence its gap closure has been associated with an instability towards a Wigner crystal~\cite{girvinCollectiveExcitationGapFractional1985, girvinMagnetorotonTheoryCollective1986, kamillaExcitonicInstabilityTermination1997}. Similarly, the collective modes of FCIs play a major role in our understanding of FCI stability.
Second, magnetorotons are intriguing on their own. In the long-wavelength limit, the excitation is a spin-2 boson corresponding to a chiral graviton~\cite{yangModelWaveFunctions2012, golkarSpectralSumRules2016, liangEvidenceChiralGraviton2024}. Its study opens new avenues for understanding the interplay of correlation and geometry in strongly interacting quantum systems.

Recent theoretical works have found excitation branches similar to the FQH magnetorotons in FCIs~\cite{kousaTheoryMagnetorotonBands2025a, paulShiningLightCollective2025, shiEffectsBerryCurvature2026, wolfIntrabandCollectiveExcitations2025,
wangDynamicsLifetimeGeometric2025, shenMagnetorotonsMoireFractional2026}.
Nonetheless, these studies considered systems close to the ideal setting of the lowest Landau level (LLL). These include a perturbed or modulated LLL~\cite{kousaTheoryMagnetorotonBands2025a, paulShiningLightCollective2025} and other bands that satisfy the trace condition~\cite{shiEffectsBerryCurvature2026, wolfIntrabandCollectiveExcitations2025, wangDynamicsLifetimeGeometric2025}, allowing their mapping to curved space Landau levels~\cite{claassenPositionMomentumDualityFractional2015, wangExactLandauLevel2021, wangOriginModelFractional2023, estienneIdealChernBands2023}. The latter have proven very successful in studies of twisted transition-metal dichalcogenide homobilayers~\cite{shiAdiabaticApproximationAharonovCasher2024a}. In this sense, the twisted MoTe$_2$ magnetorotons of Ref.~\cite{shenMagnetorotonsMoireFractional2026} are also in the LLL regime.
It remains unclear which aspects of magnetoroton phenomenology are universal consequences of fractionalization and which rely on LLL physics.
Addressing this problem requires a framework that remains applicable away from the LLL regime.
However, analytical methods are lacking; most studies of FCIs away from the LLL limit rely heavily on numerical techniques.

Coupled-wire constructions (CWCs) often allow analytical treatment of interacting systems and have been highly successful in describing integer and fractional quantum Hall states \cite{sondhiSlidingPhasesMagnetic2001a, kaneFractionalQuantumHall2002, teoLuttingerLiquidNonAbelian2014a, kanePairingLuttingerLiquids2017}, doing so in an anisotropic setting that strongly deviates from the rotationally symmetric Landau level. Similar constructions exist for both integer and fractional anomalous quantum Hall effects~\cite{klinovajaIntegerFractionalQuantum2015, hsuGeneralScatteringElectronic2023, michenAdiabaticPreparationFractional2023, shavitQuantumGeometryStabilization2024}
as well as more exotic fractional states of matter~\cite{mengCoupledwireConstructionChiral2015a, sagiNonAbelianTopologicalInsulators2014,sagiImprintTopologicalDegeneracy2015}.
The model introduced in Ref.~\cite{shavitQuantumGeometryStabilization2024}
may be tuned to deviate from the trace condition, a key property of the LLL,
destabilizing the FCI phase in accordance with the highly successful geometric stability hypothesis (GSH)~\cite{liuRecentDevelopmentsFractional2024,yuQuantumGeometryQuantum2025,royBandGeometryFractional2014,emanuelUnifyingFrameworkFractional2025,jacksonGeometricStabilityTopological2015,ledwithVortexabilityUnifyingCriterion2023, canoIdealQuantumGeometry2026}. This allows the departure from the LLL setting not only in isotropy but also in quantum geometry.
The success of CWCs in describing a variety of strongly correlated and topological phases positions them as a promising tool for studying collective excitations of such systems,
and doing so away from the LLL regime,
noting that
magnetorotons have been identified in quasi-1D systems \cite{kumarNeutralExcitationsQuantum2022}.

In this work, we employ a CWC to study neutral collective modes of fractional states tuned away from the LLL limit, with the aim of elucidating which of their properties are universal to fractional states.
Our central result is a framework for collective excitations in the strong-coupling regime of coupled-wire constructions, which are shown to emerge as solitons and breathers of effective sine-Gordon (sG) theories.

We apply our approach to a CWC model for the FQHE~\cite{kaneFractionalQuantumHall2002}. Studying neutral collective modes, we recover many of the phenomenological features of the magnetoroton branch.
We proceed to apply our method to the FCI model of Ref.~\cite{shavitQuantumGeometryStabilization2024}, and study the effects of quantum geometry on the magnetoroton gap. 
We show that the gap decreases with $\langle \text{Tr} g_{\boldsymbol{k}} \rangle$, $g_{\boldsymbol{k}}$ being the Fubini–Study metric, thus connecting the key prediction of the GSH to the magnetoroton gap closure away from the ideal limit.
In addition, we find that closer-to-ideal quantum geometry may stabilize the long-wavelength chiral graviton excitations.
The emergence of magnetoroton phenomenology within a CWC suggests that magnetorotons do not rely on Landau-level physics for their existence, although LLL-like quantum geometry enhances their stability.

In this paper, we focus on the Laughlin series~\cite{laughlinAnomalousQuantumHall1983} and describe our construction explicitly for its collective modes. Nonetheless, the appearance of effective sG models is not unique to it.
A key advantage of CWCs is their treatment of interactions, which are tractable in the underlying Tomonaga-Luttinger Hamiltonians and allow the analysis of a variety of interacting quantum states of matter on the same footing. This enables the generalization of our results to other filling fractions as well as more exotic fractional states of matter, such as fractional topological insulators and non-Abelian FQH states.

\begin{figure}[!t]
    \centering
    \includegraphics[width=1.0\linewidth]{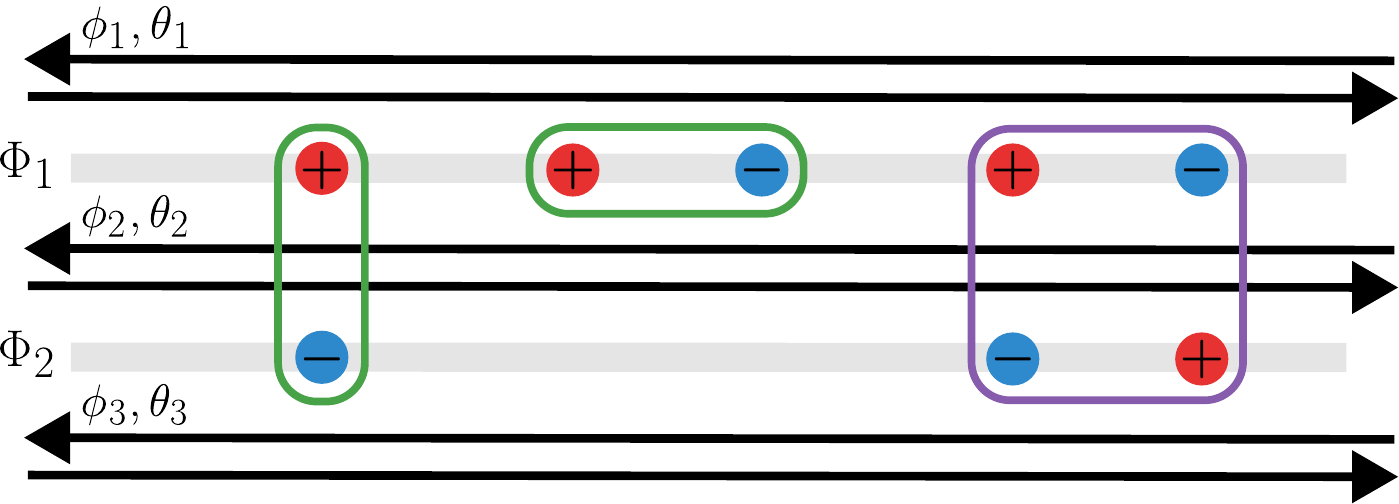}
    \caption{A schematic description of the lowest lying neutral collective modes of an FQH CWC. 
    Adjacent arrows depict left- and right-movers within a single bare wire, and are described by $\phi, \theta$ bosonic fields.
    Red (blue) dots correspond to quasiparticles (quasiholes) and are described by solitons (antisolitons) of $\Phi_j$ fields. $\Phi_j$s reside on effective wires (gray), composed of fields from two adjacent bare wires (see Eq.~\eqref{eq: variable change}). In green, dipole configurations, both inter-wire (left, vertical) as well as intra-wire (middle, horizontal). We claim that inter-wire dipoles describe the $q > q_{\text{mr}}$ part of the magnetoroton branch, with $q_{\text{mr}}$ the momentum of the magnetoroton minimum. They are described by solitons in $\mathcal{L}_{\text{2w;as;}l}$, an effective sG theory for the antisymmetric combination of two wires (see Eq.~\eqref{eq: two-wire sG params}). In purple, a quadrupole configuration corresponding to a $q=0$ excitation, described by a breather in $\mathcal{L}_{\text{2w;as;}l}$.
    }
    \label{fig: mode sketch}
\end{figure}

The rest of the paper is structured as follows.
In Section~\ref{sec: background} we provide a brief overview of the theory of magnetorotons. In Section~\ref{sec: model} we present our model of a Laughlin-series FQH system and review some key results of Ref.~\cite{kaneFractionalQuantumHall2002} on which it is based. In Section~\ref{sec: neutral exc} we construct the effective sG theories and show that their solitons correspond to the $q>q_{\text{mr}}$ modes, where $q_{\text{mr}}$ is the momentum of the magnetoroton minimum. We then discuss long-wavelength modes and their stability. 
Simply put, we construct neutral modes by combining quasiparticles and quasiholes, described by solitons and antisolitons in the model of Ref.~\cite{kaneFractionalQuantumHall2002}, as depicted pictorially in Fig.~\ref{fig: mode sketch}.
In Section~\ref{sec: stability}, we apply our method to FCIs to study the role of quantum geometry in stabilizing FCIs and their collective modes. Finally, in Section~\ref{sec: conclusions}, we summarize our results and discuss the universal aspects of collective excitations in fractional states, as well as the prospects for the applicability of our framework to the study of other collective modes.

\section{Background}
\label{sec: background}
Magnetorotons were first suggested by Girvin, MacDonald, and Platzman as collective modes of the Laughlin-series FQH states~\cite{girvinCollectiveExcitationGapFractional1985, girvinMagnetorotonTheoryCollective1986}, building upon Feynman's theory of superfluid helium excitations, the phonon-roton branch.
They proposed the single-mode approximation (SMA) wavefunction, $|\Psi_{\boldsymbol{q}}\rangle = \bar \rho_{\boldsymbol{q}} |\Psi_{\text{GS}}\rangle$, where $\Psi_{\text{GS}}$ is the ground state and $\bar \rho_{\boldsymbol{q}}$ the density operator projected to the LLL.
The mode achieves its energy minimum at a finite $q_{\text{mr}} \approx \sqrt{2\pi \nu} / \ell_B$, which may be understood as the response of the state at its characteristic length scale, the interparticle distance.
Hence, the gap closure has been associated with the transition into a Wigner crystal at a similar lattice scale.
The SMA works reasonably well for small momenta, and exceptionally well near $q_{\text{mr}}$, however, it breaks down at $q \gg q_{\text{mr}}$.

Several theories have tried to explain the nature of this collective mode. Within composite fermion theory, for example, partially filled Landau levels translate into filled ``$\Lambda$-levels'' of composite fermions~\cite{jainCompositeFermions2007}. The lowest-lying neutral modes of such systems are $\Lambda$-level excitons~\cite{kamillaExcitonicInstabilityTermination1997} carrying a nonzero dipole moment. In the quasiparticle picture, it may be understood as a bound pair of a quasiparticle and a quasihole. 
These dipoles correspond to the $q > q_{\text{mr}}$ part of the magnetoroton branch, as explained by the Chern-Simons-Landau-Ginzburg (CSLG) theory of the FQHE~\cite{zhangCHERNSIMONSLANDAU1992}. 

Projected to the LLL, position operators of quasiparticles and quasiholes satisfy $\left[x_i ^\mu, x_j ^\nu\right] \propto i\text{sign}\left(Q\right)\epsilon^{\mu\nu} \delta_{ij}$, where $i,j$ denote their quasiparticle indices, $\mu,\nu$ spatial directions and $Q$ is the quasiparticle/hole charge. Namely, a position operator acts as the conjugate momentum in the perpendicular direction. From these commutation relations, it is straightforward to verify that the relative coordinate of a dipole is proportional to its perpendicular center-of-mass momentum. Thus, the $q\to\infty$ limit corresponds to an infinitely separated dipole, and we may identify $\omega_{q\to\infty} = M_{\text{qp}} + M_{\text{qh}}$, where $M_{\text{qp}}$ ($M_{\text{qh}}$) is the quasiparticle (hole) mass. This explains the SMA's failure at large momenta: electronic density modulation cannot describe the effectively free Laughlin quasiparticles.

In the CSLG theory, the quasiparticles (holes) appear as vortices (anti-vortices) and have a finite size. Therefore, the dipole description breaks down once the dipole separation is comparable to the vortex size, which happens around $q\approx q_{\text{mr}}$. At $q=0$, Kohn's theorem forbids dipole fluctuations~\cite{kohnCyclotronResonanceHaasvan1961, zhangCHERNSIMONSLANDAU1992}, and the mode is described by a quasiparticle quadrupole, or, equivalently, a bound pair of magnetorotons~\cite{parkTwoRotonBoundState2000a, ghoshModelingTwoRotonBound2001}. The $0 < q < q_{\text{mr}}$ part of the branch continuously interpolates between the dipole and quadrupole descriptions.
In the long-wavelength limit, the lowest neutral collective mode is a spin-2 boson~\cite{yangModelWaveFunctions2012, golkarSpectralSumRules2016}, as demonstrated experimentally~\cite{liangEvidenceChiralGraviton2024}. It may be interpreted as a chiral graviton stemming from the fluctuations of an intrinsic metric defining the Laughlin state~\cite{haldaneGeometricalDescriptionFractional2011}.

\section{Model}
\label{sec: model}

We consider an array of coupled wires, each described in the infrared (IR) by a Tomonaga-Luttinger Hamiltonian. We include inter-wire coupling terms that do not open a gap and respect continuous translation symmetry along the wire, discrete translation symmetry in the perpendicular axis, and $C_{2z}$ symmetry, expected in the presence of a magnetic field.
In its bosonized form, the Hamiltonian is given by
\begin{widetext}
\begin{equation}
    \label{eq: bare Hamiltonian}
    \mathcal{H}_0 = \frac{v}{2\pi} \sum_j \int dx 
        \left(
        \frac{1}{\kappa} \left(\partial_x \phi_j\right)^2 + \kappa \left(\partial_x \theta_j\right)^2
        + \sum_{l > 0} \left(
        V_{\phi\phi;l}\partial_{x} \phi_{j} \partial_x \phi _{j+l}
        + V_{\theta\theta;l}\partial_x \theta_j \partial_x \theta_{j+l}
        + V_{\theta\phi;l} \left(\partial_x \theta_j \partial_x \phi_{j+l} - \partial_x\phi_j \partial_x \theta_{j+l} \right)
        \right)\right).
\end{equation}
\end{widetext}
Here, $\partial_x \phi_j \left(x\right) = -\pi \rho_j\left(x\right)$, with $\rho_j$ the density at wire $j$. $\theta$ operators satisfy the canonical commutation relations 
$\left[
    \phi_j\left(x\right), \partial_x \theta_k\left(x'\right)
\right] = i\pi \delta_{jk} \delta \left(x' - x\right)$.
We refer the reader to Ref.~\cite{giamarchiQuantumPhysicsOne2003} for a definition in terms of the fermionic operators and an in-depth review of bosonization.

Gaps may be opened by the inclusion of large momentum scattering operators. In the presence of a magnetic field, at $1/m$ filling, the Laughlin operator
\begin{equation}
    \label{eq: LaughlinOpr}
    \mathcal{O}_{\text{Laughlin}} = \sum_j \int dx \cos \left(m\left(\phi_j + \phi_{j+1}\right) -\theta_{j+1} + \theta_j\right)
\end{equation}
conserves momentum and stabilizes an FQH state (or an integer quantum Hall state for $m=1$)~\cite{sondhiSlidingPhasesMagnetic2001a, kaneFractionalQuantumHall2002}. In all that follows, we consider a system that is deep in the FQH state and is well described by the Hamiltonian $\mathcal{H} = \mathcal{H}_0 - g\mathcal{O}_{\text{Laughlin}}$, neglecting all other
competing gap-opening operators, such as those describing charge density waves (CDWs).
The lowest-lying charged excitations in the system are composed of solitons and antisolitons, in which the argument of the cosine shifts by $2\pi$ and $-2\pi$, respectively. Such excitations carry a charge $\pm 1/m$ and are identified as the Laughlin quasiparticles and quasiholes~\cite{kaneFractionalQuantumHall2002}.

It is useful to work with variables natural to the Laughlin operator,
\begin{gather}
    \label{eq: variable change}
    \begin{aligned}
        \Phi_j & = \frac{1}{\sqrt{8}} \left(m\left(\phi_j + \phi_{j+1}\right) - \theta_{j+1} + \theta_j\right)
        \\ \Theta_j & = \frac{1}{\sqrt{2}} \left(\frac{1}{m}\left(\theta_j + \theta_{j+1}\right) - \phi_{j+1} + \phi_j\right)
    \end{aligned}
\end{gather}
such that $\mathcal{O}_{\text{Laughlin}} = \sum_j \int dx \cos \left(\sqrt{8}\Phi_j\right)$. It is straightforward to verify that these operators satisfy the canonical Tomonaga-Luttinger commutation relations.
Henceforth, we refer to the systems described by $\Phi_j$ and $\Theta_j$ as wires, and to those described by $\phi_j$ and $\theta_j$ as bare wires.
We note that the resulting Hamiltonian is quadratic in the $\Theta$ variables, allowing us to integrate them out and arrive at the following Lagrangian density for $\Phi$,

\begin{widetext}
\begin{gather}
    \label{eq: Phi lagrangian}
    \begin{aligned}
        \mathcal{L}_{\Phi} & = \mathcal{L}_{\Phi0} -g \sum_j \cos \left(\sqrt{8} \Phi _j\right)
        \\ \mathcal{L}_{\Phi0} & = \frac{1}{2\pi}\sum_{j}
            \left(
                \left(\frac{1}{v}J_{-;0}\left(\partial_{\tau}\Phi_{j}\right)^{2}+vJ_{+;0}\left(\partial_{x}\Phi_{j}\right)^{2}\right)
                +2\sum_{l>0}\left(\frac{1}{v}J_{-;l}\partial_{\tau}\Phi_{j}\partial_{\tau}\Phi_{j+l}+vJ_{+;l}\partial_{x}\Phi_{j}\partial_{x}\Phi_{j+l}\right)
            \right),
    \end{aligned}
\end{gather}
\end{widetext}
where 
\begin{equation}
    \label{eq: J definition}
    J_{\pm;l} = \int_0 ^{2\pi} \frac{dq}{2\pi} \left(\frac{u_q}{v}\right) ^{\pm 1} K_q ^{-1}\cos\left(ql\right).
\end{equation}
Here $K_{q}$ and $u_q$ are the Luttinger parameters for $\mathcal{L}_{\Phi0}$ in the Fourier basis (explicit expressions appear in Appendix~\ref{appx: Luttinger parameters}).
Our model takes the form of infinitely many coupled sG models. In the strong coupling limit, its ground state consists of constant $\Phi_j$, pinned to cosine minima. We proceed to discuss its neutral excitations.

\section{Neutral excitations}
\label{sec: neutral exc}

\subsection{The dilute soliton gas approximation}
\label{subsec: dilute}
Neglecting inter-wire coupling ($J_{\pm;l>0} = 0$), the low-lying excitations are solitons and antisolitons that exist in each wire independently of the others (see Appendix~\ref{appx: sine Gordon} for a discussion of the sG model and its excitations).
At small inter-wire coupling ($J_{\pm;l>0} / J_{\pm;0} \ll 1$), we expect this does not change by much, except for a dressing by interactions with other wires.
Since solitons and antisolitons are charged, neutral excitations are described by a combination of solitons and antisolitons whose charges sum to zero.

Due to the energetic cost of creating a soliton, the lowest-lying excitations are composed of relatively few solitons. The minimal example consists of a bound soliton--antisoliton pair. Its constituents may reside either on the same wire, forming a breather, or on two different wires. This is in accordance with the identification of the magnetoroton as a quasiparticle-quasihole dipole \cite{zhangCHERNSIMONSLANDAU1992}, as will be elaborated later on.

We expect that in wires that do not participate in the collective mode, the chances of observing a soliton are slim. Namely, solitons are dilute.
We refer to non-participating wires as idle and approximate their confining potential as $1-\cos \left(\sqrt{8} \Phi_j\right) \approx 4 \Phi_j ^2$, which would allow us to integrate them out and see how solitons are dressed.
As shown in Appendix~\ref{appx: dressing}, soliton dressing yields $\mathcal{O} \left(\left(J_{\pm;l>0} / J_{\pm;0}\right)^2\right)$ contributions, which are subleading. Hence, for the sake of simplicity, we neglect soliton dressing, keeping in mind that since CWCs fail to provide quantitative predictions, higher-order corrections are of little value.

We may now obtain a leading-order expression for the quasiparticle mass, the soliton mass on a single wire.
Omitting all wires but one, $\mathcal{L}_{\Phi}$ takes the form of an sG model with the parameters
\begin{equation}
    \label{eq: 1w sG parameters}
    K_{\text{1w}} = \frac{1}{\sqrt{J_{+;0} J_{-;0}}}; 
    \quad
    v_{\text{1w}} = v\sqrt{\frac{J_{+;0}}{J_{-;0}}};
    \quad
    g_{\text{1w}} = g,
\end{equation}
to which we refer as $\mathcal{L}_{\text{1w}}$. $\mathcal{L}_{\Phi}$ is invariant under $\Phi_j\to-\Phi_j$, which exchanges solitons and antisolitons. The quasiparticle and quasihole masses are therefore equal, and given by
\begin{equation}
    \label{eq: quasiparticle mass}
    M_{\text{qp}} = 2 \sqrt{\frac{2}{\pi} gv J_{+;0}} \left(1 - K_\text{1w}\right).
\end{equation}
The $1 - K_\text{1w}$ factor, commonly known as the Dashen-Hasslacher-Neveu (DHN) shift, is a correction due to quantum fluctuations, computed in the semiclassical approximation~\cite{dashenParticleSpectrumModel1975, rajaraman1982solitons}.

Our assumption that the Laughlin operator flows to strong coupling implies $K_{\text{1w}} < 1$~\cite{strong_coupling}, which may be attained by engineering the inter-wire interactions $V_{\phi\phi}$, $V_{\theta\theta}$ and $V_{\theta\phi}$~\cite{kaneFractionalQuantumHall2002}.
The $K_{\text{1w}} < 1$ region of the phase diagram may be divided into two distinct regimes.
For $K_{\text{1w}} < 1/2$, solitons and antisolitons attract and may form breathers, intra-wire bound states.
For $K_{\text{1w}} > 1/2$, solitons and antisolitons cannot bind, the entire intra-wire excitation spectrum is comprised of solitons, and neutral modes must be supported on at least two wires.
We first consider the case $K_{\text{1w}} > 1/2$ and defer the discussion of single-wire breathers to subsection~\ref{subsec: breathers}.

We proceed as follows. In subsection \ref{subsec: two wires}, we find the lowest-energy configurations and show how they fit within the magnetoroton branch.
We expect these to be dipoles, which, in the absence of breathers, must be supported on two wires.
In subsection \ref{subsec: q=0}, we discuss the long-wavelength part of the excitation branch.

\subsection{Quasiparticle dipoles}
\label{subsec: two wires}
Consider two wires, at a distance $l$ from one another, which we label by $\uparrow$ and $\downarrow$. Omitting all other wires, we arrive at the effective Lagrangian density,

\begin{widetext}
\begin{gather}
    \label{eq: two-wire-lagrangian}
    \begin{aligned}
        \mathcal{L}_{\text{2w};l} = & \sum_{\eta = \uparrow \downarrow}
        \left(
            \frac{1}{2\pi v}J_{-;0} \left(\partial_\tau \Phi_\eta\right)^2
            + \frac{v}{2\pi}J_{+;0} \left(\partial_x \Phi_\eta\right)^2
            - g \cos \left(\sqrt{8} \Phi_\eta\right)
        \right)
        +
        \left(
            \frac{1}{\pi v} J_{-;l} \partial_\tau \Phi_\uparrow \partial_\tau \Phi_\downarrow
            + \frac{v}{\pi}J_{+;l} \partial_x\Phi_\uparrow \partial_x \Phi_\downarrow
        \right),
    \end{aligned}
\end{gather}
which takes the form of two sG models, with quadratic coupling. We look for a neutral mode with a soliton in one wire and an antisoliton in the other.
The inter-wire interaction is contact-like in the $x$ coordinate, and we expect it to be attractive between quasiparticles and quasiholes. Hence, the minimal energy configuration would be of an aligned quasiparticle and quasihole, $\Phi_\downarrow = - \Phi_\uparrow$.
Pinning the symmetric channel, which we expect to be stiffer than the antisymmetric one, we obtain the Lagrangian $\mathcal{L}_{\text{2w;as;}l}\left[\Phi\right]$ which takes the form
of an sG model with the parameters
\begin{equation}
    \label{eq: two-wire sG params}
     K_{\text{2w;as;}l} = \frac{1}{2\sqrt{\left(J_{+;0} - J_{+;l}\right)\left(J_{-;0} - J_{-;l}\right)}}; \quad\quad
    v_{\text{2w;as;}l} = v\sqrt{\frac{J_{+;0} - J_{+;l}}{J_{-;0} - J_{-;l}}}; \quad\quad
    g_{\text{2w;as;}l} = 2g,
\end{equation}
\end{widetext}
where ``as'' stands for antisymmetric. It is straightforward to verify from Eq.~\eqref{eq: J definition} that $|J_{\pm;l}| < J_{\pm;0}$, so the parameters are well defined.

A soliton in this model corresponds to a quasiparticle-quasihole dipole, as depicted on the left of Fig.~\ref{fig: mode sketch}.
The classical soliton mass may be simply extracted from $\mathcal{L}_{\text{2w;as;}l}$. However, quantum fluctuations in the symmetric channel contribute to the DHN shift, leading to the mass
\begin{equation}
    \label{eq: two-wire excitation energy}
    M_{\text{2w;as;}l} = 4\sqrt{\frac{2}{\pi}gv\left(J_{+;0} - J_{+;l}\right)} \left(1 - K_{\text{2w;as;}l}\left(1 + \delta_l\right)\right).
\end{equation}
The correction to the DHN shift $\delta_l$ is calculated in Appendix~\ref{appx: dhn} using the semiclassical approximation with the soliton--antisoliton separation treated as a collective coordinate. 

As a soliton, the dipole has topological charge $1$ in $\mathcal{L}_{\text{2w;as;}l}$, yet is electrically neutral. We stress that physical electric charge and topological charge in sG theories generally differ.
For $l\to\infty$, $J_{\pm;l} \to 0$, and thus for sufficiently large wire separation $l$, $M_{\text{2w;as;}l} \approx 2M_{\text{qp}}$ as expected~\cite{large_l_note}. The magnetoroton gap is estimated by $E_\text{mr} = \min_l M_{\text{2w;as;}l}$.
We expect solitons and antisolitons to attract such that the minimum is attained for the minimal separation $l=1$, namely, $J_{+;1} > 0$ and $J_{+;1} \ge J_{+;l}$ for $l>1$.
However, these conditions are not guaranteed.

\begin{figure}[!t]
    \centering
    \includegraphics[width=1.0\linewidth]{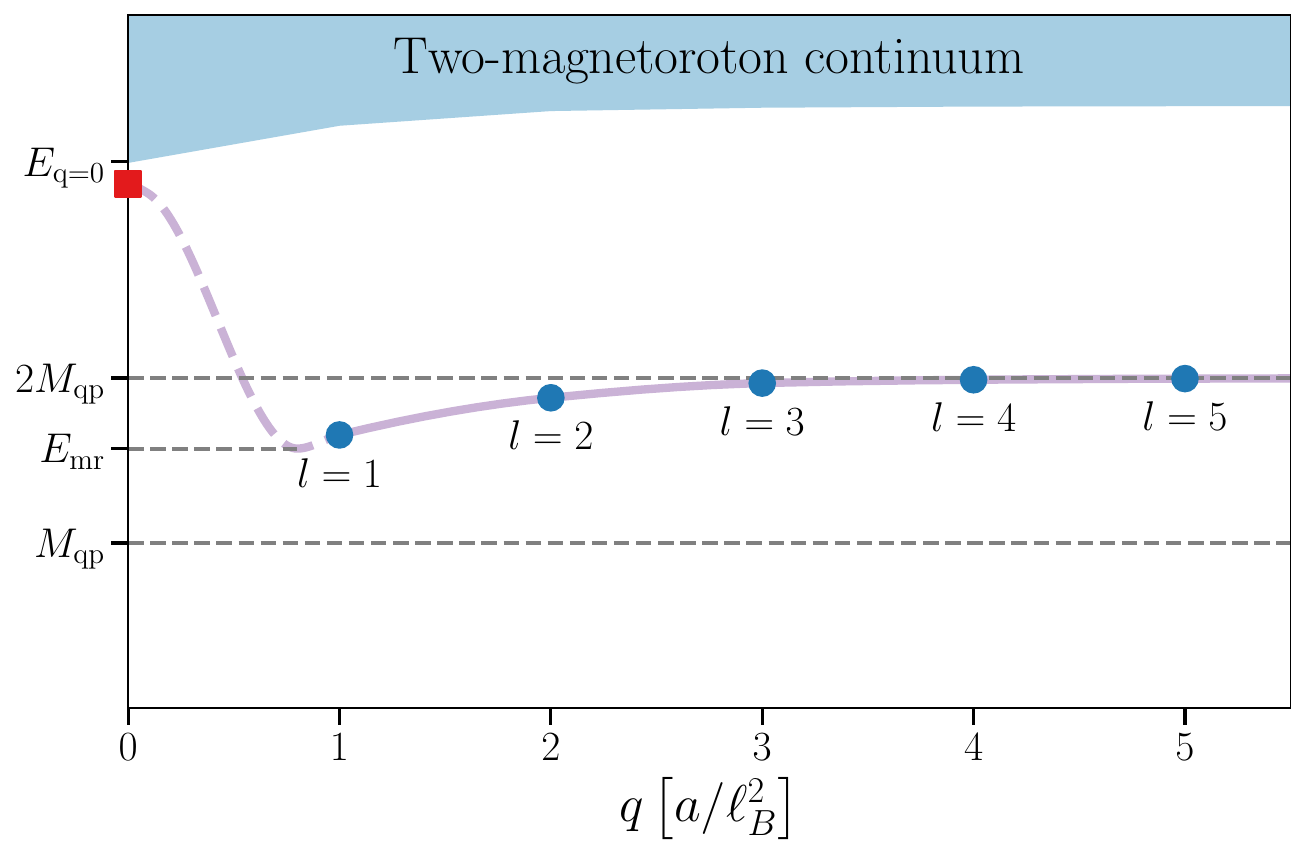}
    \caption{Dispersion of the neutral collective modes.
    Blue dots on the right, dipole energies $M_{\text{2w;as;}l}$ for different dipole separations $l$. 
    Red square, the quadrupole ($\mathcal{L}_{\text{2w;as;}l}$ breather) energy $E_{q=0} \equiv 2M_{\text{2w;as;}l=1} \sin \left( \frac{\pi}{2}\frac{K_{\text{2w;as;}l=1}\left(1 + \delta_{l=1}\right)}{1 - K_{\text{2w;as;}l=1}\left(1 + \delta_{l=1}\right)}\right)$. For this choice of parameters, it lies below the two-magnetoroton continuum at $2E_{\text{mr}}$ and is hence long-lived.
    In lilac, a sketch of the magnetoroton branch. The solid line ($q > q_{\text{mr}}$) denotes the regime where a quasiparticle-quasihole dipole well describes the excitation. The dashed line ($q < q_{\text{mr}}$) denotes the regime in which the excitation interpolates between a dipole and a quadrupole, which is unattainable within our effective sG approach.
    The parameters are given by a $V_{\theta \phi} = 0.85$ line cut of Fig.~\ref{fig: FCI magnetorotons} (bottom panel).}
    \label{fig: branch sketch}
\end{figure}

The correspondence of excitations in our CWC to those of the CSLG is immediate.
The neutral modes in our wire construction are precisely quasiparticle-quasihole dipoles, which we may identify, following the CSLG theory, with the large momentum part of the excitation branch $q > q_\text{mr}$, where the momentum is proportional to the dipole separation, as depicted in Fig.~\ref{fig: branch sketch}. 
In the CWC, position-momentum duality manifests as wire-dependent momentum shifts, which ensure the momentum conservation of $\mathcal{O}_{\text{Laughlin}}$ and allow its addition to the Hamiltonian~\cite{kaneFractionalQuantumHall2002}.
Unlike in the continuum, quasiparticles and quasiholes cannot overlap since they reside on different wires. Therefore, we do not observe a smooth minimum in the excitation branch, and cannot obtain $q < q_\text{mr}$ excitations within the effective-sG framework. 
There is, however, a natural candidate for the $q=0$ excitation.

\subsection{$q = 0$}
\label{subsec: q=0}
Our assignment of a finite momentum 
to inter-wire soliton--antisoliton dipoles implies that an excitation at $q = 0$ must have more than two constituents. For a neutral mode, the lowest number of which would now be two solitons and two antisolitons, and can be thought of as a bound pair of dipoles. The momenta of the dipoles need to sum to zero, and therefore they should have the same separation, but with an opposite sign (the soliton and antisoliton interchanged). Equivalently, we choose the dipoles to be $C_{2z}$ partners of one another since $C_{2z}$ takes $\boldsymbol{q} \to -\boldsymbol{q}$.

Our construction yields an anisotropic quadrupole with a net dipole moment of zero (Fig.~\ref{fig: mode sketch}, right). A $\pi$-rotation acts as $\Phi_j \left(x\right) \to -\Phi_{-j} \left(-x\right)$ (see Appendix~\ref{appx: Luttinger parameters} and Eq.~\eqref{eq: variable change}), so a soliton is $C_{2z}$-invariant.
Therefore, the quadrupole configuration is invariant under a rotation by $\pi$, implying it carries an even spin, in agreement with the expectation of a spin-2 boson \cite{golkarSpectralSumRules2016}. To distinguish higher spins, one would need to check transformations under rotation by $\theta < \pi$, a meaningless operation in our anisotropic CWC.

The dipoles may always be chosen such that they do not interact, placing the upper bound for the lowest energy state $E_{q=0} \le 2 \min_l M_{\text{2w;as;}l} = 2E_{\text{mr}}$.
However, such a mode would easily decay into the two-magnetoroton continuum and is thus unstable.
Dipoles placed on the same set of wires may bind, forming a breather with an energy strictly lower than $2M_{\text{2w;as;}l}$. Typically, the condition for breather formation in an sG theory is $K_{\text{2w;as;}l} < 1 /2$. However, quantum fluctuations in the symmetric channel which alter the DHN shift may modify the breather-formation condition as well. Here and below, we consider the stricter heuristic condition $K_{\text{2w;as;}l}\left(1 + \delta_l\right) < 1 /2$ for breather formation.
Since $M_{\text{2w;as;}l}$ is typically minimal for $l=1$, the energy of $l > 1$ breathers might still exceed $2E_{\text{mr}}$, lie within the two-magnetoroton continuum and become unstable.

The system may be tuned such that the long-wavelength modes are either long-lived or short-lived.
Whether these excitations are long-lived in FCIs,
where the continuous rotation symmetry protecting the chiral graviton is absent,
has been a subject of debate~\cite{longChiralGravitonModes2026, wangDynamicsLifetimeGeometric2025}. Our results provide a convenient way to identify the parameters determining the stability of the long-wavelength mode.

\subsection{Intra-wire dipoles}
\label{subsec: breathers}
Lastly, we return to the discussion of intra-wire dipoles (Fig.~\ref{fig: mode sketch}, middle), which appear if $K_{\text{1w}} < 1/2$~\cite{breather_modes}.
Naively, using the same analysis, we identify them as $q=0$, odd-spin particles. 
However, this raises a physical difficulty. In subsection~\ref{subsec: two wires},
we identified the separation of a dipole in the direction perpendicular to the wires (a classical label) with its momentum along the wires, using the wire momentum shift induced by a magnetic field. This relation is non-reciprocal: an intra-wire dipole is pinned to a single wire and its separation along the wire is not tied to any transverse momentum. In a Landau level, position-momentum duality stems from the non-commutativity of the projected position coordinates, which is reciprocal: the separation of a dipole in any direction is proportional to the center-of-mass conjugate momentum in the perpendicular direction.
Therefore, we expect that realistically, intra-wire breathers do not form a distinct $\boldsymbol{q} = 0$ mode. In all that follows, we discard them, focusing on the rotation-equivalent but better-controlled inter-wire dipoles.

\section{Fractional state stability}
\label{sec: stability}

Having identified key features of the magnetoroton branch in a CWC model for the FQH, we turn our attention to FCIs and apply our method to the model introduced in Ref.~\cite{shavitQuantumGeometryStabilization2024}.
Here, we present the 
relevant findings from that work, with an emphasis on the relation of quantum geometry to FCI stability and to the parameters of Eq.~\eqref{eq: bare Hamiltonian}. We refer the reader to Appendix~\ref{appx: FCI} for details and to the original paper for explicit expressions and complete derivations.

Ref.~\cite{shavitQuantumGeometryStabilization2024} presents a novel ordering tendency, the ``anti-FCI'' (aFCI), which is an FCI of opposite chirality. Within the CWC framework, an FCI operator $\mathcal{O}_{\text{FCI}}$ is simply $\mathcal{O}_{\text{Laughlin}}$ of Eq.~\eqref{eq: LaughlinOpr}, and an aFCI operator $\mathcal{O}_{\text{aFCI}}$ takes the same form with $m\to -m$.
The aFCI does not flow to strong coupling~\cite{tautology_statement}, yet it hinders FCI formation and enlarges the CDW region of the phase diagram.
This can be straightforwardly understood from the model's RG equations (Eq.~\eqref{eq: RG flow}). During its flow, $\mathcal{O}_{\text{FCI}}$ generates chiral interactions ($\propto V_{\theta\phi}$ of Eq.~\eqref{eq: bare Hamiltonian}), reducing its own dimension. The operator $\mathcal{O}_{\text{aFCI}}$ generates chiral interactions as well, but with the opposite sign, increasing the dimension of $\mathcal{O}_{\text{FCI}}$ and delaying the latter's flow to strong coupling.

\begin{figure}[!t]
    \centering
    \includegraphics[width=1.0\linewidth]{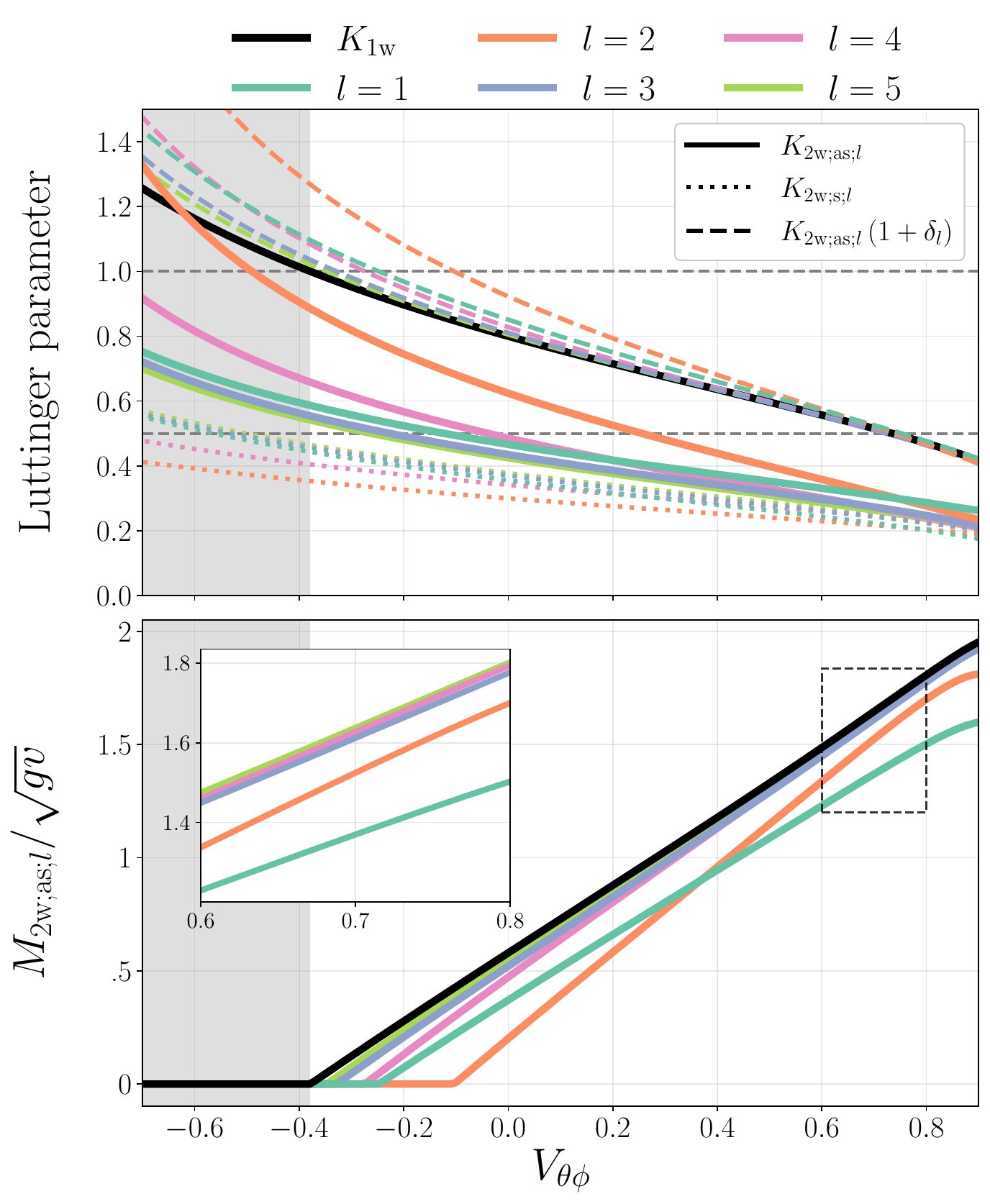}
    \caption{
        Top: Solid lines, Luttinger parameters $K_{\text{1w}}$ (Eq.~\eqref{eq: 1w sG parameters}) and $K_{\text{2w;as;}l}$ (Eq.~\eqref{eq: two-wire sG params}) with $l=1,\dots,5$ vs. $V_{\theta\phi}$ (the scale of $V_{\theta\phi;l}$ interactions in Eq.~\eqref{eq: bare Hamiltonian}).
        Dotted lines, Luttinger parameters of the corresponding symmetric mode $K_{\text{2w;s;}l}$ (same as Eq.~\eqref{eq: two-wire sG params}, taking $J_{\pm;l} \to - J_{\pm;l}$). $K_{\text{2w;s;}l} < K_{\text{2w;as;}l}$ in accordance with our expectation of a stiffer symmetric mode.
        Dashed lines, DHN shifts $K_{\text{2w;as;}l}\left(1 + \delta_l\right)$ (Eq.~\eqref{eq: two-wire excitation energy}). When the shifts are below $1/2$, quadrupoles may form.
        Bottom: Corresponding dipole energies $M_{\text{2w;as;}l}$ (Eq.~\eqref{eq: two-wire excitation energy}), and $2M_{\text{qp}}$ (Eq.~\eqref{eq: quasiparticle mass}, black).
        Inset, a zoom-in on the dipole energies in the $V_{\theta\phi} \in \left[0.6, 0.8\right]$ region (dashed square), showing $l > 2$ excitations where $M_{\text{2w;as;}l} \approx 2M_{\text{qp}}$.
        In gray, regions where $K_{\text{1w}} > 1$, the FCI operator is irrelevant and our theory is inapplicable.
        The parameters are
        $\kappa = 0.21$, $V_{\phi\phi} = 5.0$, $V_{\theta\theta} = -0.15$, $\lambda = 0.15$ and $m=3$.}
    \label{fig: FCI magnetorotons}
\end{figure}

In the model of Ref.~\cite{shavitQuantumGeometryStabilization2024}, the difference of the UV values of the FCI and aFCI coupling constants, $g_{\text{FCI}} - g_{\text{aFCI}}$,
is set by the quantum metric of the underlying lattice model: $g_{\text{FCI}} \approx g_{\text{aFCI}}$ corresponds to large values of $\langle \text{Tr} g_{\boldsymbol{k}} \rangle$ while $g_{\text{FCI}} \gg g_{\text{aFCI}}$ corresponds to small ones. Thus, the geometric stability hypothesis (GSH) is explained in terms of FCI--aFCI competition.
The above analysis further implies that the IR value of $V_{\theta\phi}$ acts as a good proxy for the UV quantum geometry: 
$\langle \text{Tr} g_{\boldsymbol{k}} \rangle$ sets the initial conditions for the FCI-aFCI competition, which determines how much $V_{\theta\phi}$ is generated during the RG flow.

We proceed to examine the effect of $V_{\theta\phi}$ on the magnetoroton mass and Luttinger parameters of the effective sG models. We consider an underlying model of the form of Eq.~\eqref{eq: bare Hamiltonian}, with exponentially decaying inter-wire interactions, $V_{\alpha\beta;l} = V_{\alpha\beta} \lambda^{l-1}$ ($\alpha\beta \in \{\phi\phi, \theta\theta, \theta\phi\}$, $\lambda < 1)$.
Our results are presented in Fig.~\ref{fig: FCI magnetorotons}. 
Sufficiently large $V_{\theta\phi}$ is required to get $K_\text{1w} < 1$ and stabilize the fractional state.
Minimal energies are attained for a dipole separation $l=1,2$. The gap increases with $V_{\theta\phi}$ and hence, via the proxy argument above, it decreases with $\langle \text{Tr} g_{\boldsymbol{k}} \rangle$, showing a connection between the GSH and the magnetoroton gap.
The monotonic decrease of the Luttinger parameters $K_{\text{2w;as;}l}$ with $V_{\theta\phi}$ suggests that quantum geometry helps stabilize the quadrupole mode.
For sufficiently large $V_{\theta\phi}$, $K_{\text{2w;as;}l=1}\left(1 + \delta_{l=1}\right) < 1/2$, implying the existence of a sharp breather mode in $\mathcal{L}_{\text{2w;as;}l=1}$: the long-wavelength quadrupole is long-lived.
Within this model, ideal quantum geometry not only stabilizes fractional states but also helps stabilize their long-wavelength neutral collective modes.

\section{Conclusions}
\label{sec: conclusions}

We developed a new method to study collective modes of CWCs in the strong-coupling limit. For a CWC with a single massive operator, the Lagrangian in Eq.~\eqref{eq: Phi lagrangian} is general. One only needs to derive the problem-specific effective-wire operators $\Phi_j$ and the coefficients $J_{\pm;l}$. To leading order in the inter-wire coupling, excitations, both charged and neutral, are well described by solitons and breathers in effective sG models and correspond to modes sketched in Fig.~\ref{fig: mode sketch}.
We employed this approach to study CWC models for the Laughlin series FQH~\cite{kaneFractionalQuantumHall2002} and an FCI~\cite{shavitQuantumGeometryStabilization2024}. 
Both differ from the LLL in their anisotropy, the latter also in its quantum geometry, prompting this study of collective modes in CWCs.

Despite the limitations of CWCs, our approach captures many key features of the magnetoroton branch. We can describe its $q > q_{\text{mr}}$ as well as its $q \approx 0$ parts, covering the same range as the isotropic CSLG theory~\cite{zhangCHERNSIMONSLANDAU1992}.
Anisotropy and discreteness prevent us from identifying the exact spin of the $q=0$ mode, which we can only tell mod 2. Even so, the quadrupole nature of the excitation suggests that the spin would remain 2 in a more isotropic scenario, as expected~\cite{golkarSpectralSumRules2016}.

The reproduction of key properties of the magnetoroton phenomenology, even under strong anisotropy, can be traced to two observations. First, if the IR theory has charged quasiparticles/quasiholes as fundamental excitations, neutral modes take the form of quasiparticle dipoles, quadrupoles, etc. Second, position-momentum duality in the presence of a magnetic field allows one to substitute dipole separations for center-of-mass momenta. These two observations hold for both the isotropic Landau level and our CWCs (the latter only in one orientation), leading to the qualitatively similar excitation spectrum. This implies FCIs may have a similar excitation spectrum, even away from the Landau-level limit.
Note that the arguments above do not rely on the carriers' fractional charge, and would hold for a neutral excitation composed of integer-charge particles. Therefore, we expect excitons, formed of electrons and holes, could display a similar magnetoroton-like spectrum in the presence of a magnetic field. This is in agreement with the obtained spectra for magnetoexcitons in filled Landau levels~\cite{kallinExcitationsFilledLandau1984a}.

Recently, it has been suggested that quasiparticles of identical charges may bind into anyon molecules~\cite{gattuMolecularAnyonsFractional2025, wangAnyonMoleculesFractional2026}. Binding is possible due to oscillations in the anyon's charge density, which contains both positive and negative regions (relative to the ground state). When negative-density regions of one anyon overlay positive-density regions of the other, they may bind even if their total charges have the same sign.
Density oscillations are missing from our soliton description of quasiparticles. This is compatible with the long-wavelength approximation underlying CWCs, which misses intricate, short-range details of the anyon wavefunctions. As a result, we do not expect our model to describe anyon molecules faithfully.

Our scheme can be straightforwardly applied to study excitations of other strongly interacting states of matter.
Replacing Eq.~\eqref{eq: variable change} with an appropriate variable change, one only needs to re-derive $J_{\pm;l}$, a matter of standard Gaussian integration.
See, for example, Eq.~(2.45) of Ref.~\cite{teoLuttingerLiquidNonAbelian2014a}, from which one may naturally define a suitable variable for second-level hierarchy FQH states.
Additional systems to which this approach could be applied include non-Abelian FQH states, topological insulators, and chiral spin liquids~\cite{teoLuttingerLiquidNonAbelian2014a,
kanePairingLuttingerLiquids2017,
sagiNonAbelianTopologicalInsulators2014, mengCoupledwireConstructionChiral2015a, sagiImprintTopologicalDegeneracy2015}.

We find the magnetoroton gap of the FCI model decreases with $\langle \text{Tr} g_{\boldsymbol{k}} \rangle$, connecting the magnetoroton gap closure with the celebrated GSH. Even though the GSH originates in an attempt to mimic the LLL~\cite{royBandGeometryFractional2014}, our work joins existing evidence that quantum geometric properties play a significant role in stabilizing FCIs even away from the LLL regime~\cite{shavitQuantumGeometryStabilization2024, emanuelUnifyingFrameworkFractional2025, andrewsStabilityFractionalChern2024}.
Furthermore, our results suggest that the stability of the $q\approx 0$ chiral graviton in FCIs may also be quantum-geometry dependent. Similar observations were made in Ref.~\cite{longChiralGravitonModes2026} using different methods. Though additional evidence is required to verify the last conclusion, it exemplifies the strength of CWCs in identifying trends with relative ease and in an intuitive manner, precisely where the application of many other techniques proves challenging.

\section*{Acknowledgments}
The authors thank Noga Bashan and Gal Shavit for illuminating discussions and the Kavli Institute for Theoretical Physics (KITP) for its hospitality. 
This research was supported by the Israel Science Foundation (ISF, Grant No. 1914/24),
ISF Quantum Science and Technology program (Grant No. 2074/19), the Minerva Stiftung and the Deutsche Forschungsgemeinschaft (DFG) via CRC/Transregio 183, (EI 519/7-1), and, in part, by grant NSF PHY-2309135 to the KITP.
PE is supported by the Chaim Mida Prize and by the CHE/PBC fellowship for Outstanding Doctorate students in Quantum Science and Technology.
The authors acknowledge the use of Anthropic's Claude Opus 5.5 to assist in drafting the code used to generate Figs.~\ref{fig: branch sketch} and~\ref{fig: FCI magnetorotons}.

\bibliography{references, notes}

\begin{thebibliography}{76}
\bibitem [{str()}]{strong_coupling}%
  \BibitemOpen
  \bibinfo {note} {The condition $K_{\text{1w}}<1$ is necessary but not sufficient. At the Gaussian fixed point, the scaling dimension of $\cos(\sqrt{8}\Phi_j)$ is $2\bar{K}$, where $\bar{K}\equiv\int(dq/2\pi)K_q$, and the Cauchy--Schwarz inequality gives $K_{\text{1w}}\le\bar{K}$.}\BibitemShut {Stop}%
\bibitem [{lar()}]{large_l_note}%
  \BibitemOpen
  \bibinfo {note} {We note that for sufficiently large $l$, dipole-wire terms $\left(\mathcal{O}\left(J_{\pm;l} / J_{\pm;0}\right)\right)$ become smaller than quadratic contributions from wires not participating in the dipole $\left(\mathcal{O}\left(\left(J_{\pm;1} / J_{\pm;0}\right)^2\right)\right)$ since the latter do not decay with $l$. The contributions from the idle wires can no longer be neglected, and our construction becomes inapplicable.}\BibitemShut {Stop}%
\bibitem [{bre()}]{breather_modes}%
  \BibitemOpen
  \bibinfo {note} {Here, and whenever breathers are discussed, we refer to the ``first'', lowest energy, breathers. Additional branches form at lower Luttinger parameters, the next one at $K < 1/3$ (see Appendix~\ref{appx: sine Gordon}). These go beyond the scope of the qualitative discussion of the lowest neutral excitations in this paper.}\BibitemShut {Stop}%
\bibitem [{tau()}]{tautology_statement}%
  \BibitemOpen
  \bibinfo {note} {By convention, if $\mathcal{O}_{\text{aFCI}}$ flows to strong coupling rather than $\mathcal{O}_{\text{FCI}}$, we swap their definitions, referring to the former as the FCI operator and to the latter as the anti-FCI operator.}\BibitemShut {Stop}%
\end{thebibliography}

\onecolumngrid
\appendix
\section{The sine-Gordon model}
\label{appx: sine Gordon}
We briefly overview key properties of the sine-Gordon (sG) model and its excitations. For an in-depth review and derivations of the results below, we refer the readers to
Refs.~\cite{cuevas-maraverSineGordonModelIts2014, rajaraman1982solitons, giamarchiQuantumPhysicsOne2003,dashenParticleSpectrumModel1975, colemanQuantumSineGordonEquation1975}. We follow the conventions of Giamarchi~\cite{giamarchiQuantumPhysicsOne2003}. 
The quantum sG model is a $\left(1+1\right)$-dimensional theory whose Lagrangian density is given by
\begin{equation}
    \mathcal{L}_{\text{sG}} \left[\varphi\right] = 
            \frac{1}{2\pi K_{\text{sG}}} \left(
            \frac{1}{v_{\text{sG}}} \left(\partial_\tau \varphi\right)^2
            + v_{\text{sG}} \left(\partial_x \varphi\right)^2
            \right)
            + g_{\text{sG}}\left(1 - \cos\left(\sqrt{8} \varphi\right)\right).
\end{equation}
For $K_\text{sG} < 1$, the cosine is relevant and opens a gap.
The equation of motion reads
\begin{equation}
    \frac{1}{v_{\text{sG}}}\partial_{\tau}^{2}\varphi
    + v_{\text{sG}}\partial_{x}^{2}\varphi
    -2\sqrt{2}\pi K_\text{sG}g_{\text{sG}}\sin\left(\sqrt{8}\varphi\right)=0
\end{equation}
and may be turned into an ordinary differential equation 
with a traveling-wave Ansatz with velocity $c$, $\varphi\left(x,\tau\right) = \varphi\left(x-ic\tau\right)$. In all that follows, we restrict ourselves to the static case $c=0$. Dynamic solutions may be attained from static ones via a boost ($\mathcal{L}_{\text{sG}}$ is Lorentz-invariant).
The static soliton solution reads
\begin{equation}
    \label{eq: sG soliton}
    \varphi_{\text{s}} = \sqrt{2}\arctan\left(e^{x/L_s}\right)
\end{equation}
and interpolates between $\sqrt{8}\varphi_s = 0$ at $x=-\infty$ and $\sqrt{8}\varphi_s = 2\pi$ at $x=\infty$. Antisolitons interpolate between $2\pi$ and $0$, and are given by $\varphi_{\text{as}} \left(x\right) = \varphi_{\text{s}} \left(-x\right)$. Here
$L_s =\left(8\pi g _\text{sG}K _\text{sG} / v_\text{sG}\right)^{-1/2}$
is the soliton width. Under the semiclassical approximation, the soliton's mass is given by
\begin{equation}
    \label{eq: sG soliton mass}
    M_{\text{sol}} = 2 \sqrt{\frac{2g_{\text{sG}} v_{\text{sG}}}{\pi K_{\text{sG}}}} \left(1-K_{\text{sG}}\right).
\end{equation}
It differs from the classical mass by a $\left(1-K_{\text{sG}}\right)$ factor, commonly known as the DHN shift~\cite{dashenParticleSpectrumModel1975, rajaraman1982solitons}. Note that the mass cited in Appendix E of Ref.~\cite{giamarchiQuantumPhysicsOne2003} (Eq.~(E43)) is the classical mass. Eq.~\eqref{eq: sG soliton mass} is given by Eq.~(7.10) of Ref.~\cite{rajaraman1982solitons}, with the notation dictionary of Ref.~\cite{giamarchiQuantumPhysicsOne2003}'s Eq.~(E41). It is also derived in Appendix~\ref{appx: dhn}.
For $K_{\text{sG}} < 1/2$, solitons and antisolitons attract, and may form bound states known as breathers, whose masses are
\begin{equation}
    \label{eq: breather}
    M_{\text{bre};n} = 2 M_{\text{sol}} \sin \left(\frac{\pi K_\text{sG} n}{2\left(1 - K_\text{sG}\right)}\right)
\end{equation}
with $1 \le n < K_\text{sG} ^{-1} - 1$ an integer. We note that both the soliton and the breather masses are proportional to the only energy scale in the system, $\sqrt{g_{\text{sG}} v_{\text{sG}}}$.

\section{Model parameters}
\label{appx: Luttinger parameters}
In this appendix, we provide additional details regarding the model parameters. We derive the Luttinger parameters and relate the parameters of Eq.~\eqref{eq: bare Hamiltonian} to those of Eq.~\eqref{eq: Phi lagrangian}.

A general coupled wire construction Hamiltonian
is given by coupled Tomonaga-Luttinger Hamiltonians.
In the presence of a magnetic field, and neglecting disorder, we restrict our discussion to couplings respecting $C_{2z}$ and translation symmetry, where $z$ is the axis perpendicular to the wire plane.
Since $\partial_x \phi \approx -\pi\left(\rho_R + \rho_L\right)$ and $\partial_x \theta \approx \pi \left(\rho_R - \rho_L\right)$, where $\rho_{R(L)}$ is the density of right (left) movers in the wire, $C_{2z}$ acts on the fields as $\partial_x \phi \to \partial_x \phi$ and $\partial_x \theta \to -\partial_x \theta$. Together with wire translation, it may be seen as exchanging two wires. These restrictions leave three allowed quadratic operators that do not open a gap,

\begin{equation*}
    \text{I)}\quad \partial_x \phi_j \partial_x \phi_{j+l} \quad\quad\quad
    \text{II)}\quad \partial_x \theta_j \partial_x \theta_{j+l} \quad\quad\quad
    \text{III)}\quad \partial_x \theta_j \partial_x \phi_{j+l} - \partial_x\phi_j \partial_x \theta_{j+l}, \quad\quad\quad
\end{equation*}
leading to the general form of $\mathcal{H}_0$ in Eq.~\eqref{eq: bare Hamiltonian}.
The first term describes inter-wire density-density interactions. The second leads to a difference in interaction between fermions of the same or opposite chirality. The last term chooses a specific chirality. It contributes to left-mover-right-mover interaction with opposite signs depending on which chirality is in the higher-indexed wire. It is generated by the Laughlin operator as shown in Appendix~\ref{appx: FCI}.
We utilize translation symmetry in the direction perpendicular to the wires to write

\begin{equation}
        \mathcal{H}_0 = \sum_{q_\perp} \frac{v_{q_\perp}}{2\pi} \int dx
        \left(
            \frac{1}{\kappa_{q_\perp}} \partial_x \phi _{q_\perp} \partial_x \phi _{-q_\perp}
            + \kappa_{q_\perp} \partial_x \theta _{q_\perp} \partial_x \theta _{-q_\perp}
            +i\eta_{q_{\perp}}\left(\partial_{x}\phi_{q_{\perp}}\partial_{x}\theta_{-q_{\perp}}-\partial_{x}\theta_{q_{\perp}}\partial_{x}\phi_{-q_{\perp}}\right)
        \right)
\end{equation}
where the Fourier transform of an operator $A$ is given by $A_j = N^{-1/2} \sum_{q_\perp} \exp\left(iq_\perp j\right) A_{q_\perp}$, with $N$ the number of wires. The Luttinger parameters are given by
\begin{gather}
    \begin{aligned}
        \frac{v_{q_\perp}}{v} & = \sqrt{
            \begin{aligned}
                & \left(1+\kappa\sum_{l>0}V_{\phi\phi;l}\cos\left(lq_{\perp}\right)\right)
                \left(1+\kappa^{-1}\sum_{l>0}V_{\theta\theta;l}\cos\left(lq_{\perp}\right)\right)
            \end{aligned}
        },
        \\
        \kappa_{q_\perp} & = \kappa\sqrt{\frac{1+\kappa^{-1}\sum_{l>0}V_{\theta\theta;l}\cos\left(lq_{\perp}\right)}{1+\kappa\sum_{l>0}V_{\phi\phi;l}\cos\left(lq_{\perp}\right)}}, \text{ and}
        \\ \eta_{q_\perp} & = \frac{v}{v_{q_{\perp}}}\sum_{l>0}V_{\theta\phi;l}\sin \left(lq_{\perp}\right).
    \end{aligned}
\end{gather}
The theory is stable as long as $\kappa_{q_\perp}, v_{q_\perp} \in \mathbb{R}^+$ and $|\eta_{q_\perp}| < 1$ for every $q_\perp$.

By performing the variable change defined in Eq.~\eqref{eq: variable change}, we arrive at a Lagrangian where all terms are quadratic in the $\Theta$ fields, which may be integrated out, yielding the Lagrangian density
\begin{equation}
    \mathcal{L}_{\Phi0} = \sum_{q_{\perp}}\frac{1}{2\pi K_{q_{\perp}}}\left(\frac{1}{u_{q_{\perp}}}\partial_{\tau}\Phi_{q_{\perp}}\partial_{\tau}\Phi_{-q_{\perp}}
    +u_{q_{\perp}}\partial_{x}\Phi_{q_{\perp}}\partial_{x}\Phi_{-q_{\perp}}\right),
\end{equation}
where
\begin{equation}
    K_{q_\perp} = \frac{m}{2}\frac{1}{\sqrt{1-\eta_{q_{\perp}}^{2}}}
            \left(
                \frac{1}{m\kappa_{q_{\perp}}}\sin^{2}\left(\frac{q_{\perp}}{2}\right)
                +m\kappa_{q_{\perp}}\cos^{2}\left(\frac{q_{\perp}}{2}\right)
                -\eta_{q_{\perp}}\sin q_{\perp}
            \right); \quad\quad\quad
    u_{q_\perp} = v_{q_{\perp}}\sqrt{1-\eta_{q_{\perp}}^{2}}.
\end{equation}
Performing the inverse Fourier transform, we get $\mathcal{L}_{\Phi 0}$ of Eq.~\eqref{eq: Phi lagrangian} and
\begin{equation}
    J_{\pm;l} = 
    \frac{2}{m}\int_0 ^{2\pi} \frac{dq}{2\pi}
    \frac{
        \left(1-\eta_{q}^{2}\right)^{\left(1\pm1\right)/2}\left(v_{q}/v\right)^{\pm1} \cos \left(lq\right)
    }{
        \frac{1}{m\kappa_{q}}\sin^{2}\left(\frac{q}{2}\right)+m\kappa_{q}\cos^{2}\left(\frac{q}{2}\right)-\eta_{q}\sin q
    }.
\end{equation}

\section{Quantum shifts in the two-wire model}
\label{appx: dhn}
When deriving $\mathcal{L}_{\text{2w;as;}l}$ from $\mathcal{L}_{\text{2w};l}$ (Eq.~\eqref{eq: two-wire-lagrangian}), we substituted $\Phi_\uparrow = -\Phi_\downarrow$. By doing so, we neglected the fluctuations of the symmetric combination $\Phi_\uparrow + \Phi_\downarrow$. 
These do, however, contribute to the DHN shift of the two-wire soliton mass, which we calculate in this appendix.
We consider a model of two coupled sG Lagrangians (Eq.~\eqref{eq: two-wire-lagrangian} takes this form),
\begin{equation}
    \label{eq: two coupled sG Lagrangian}
        \mathcal{L}_{\text{csG}} \left[\varphi_1, \varphi_2\right] = \mathcal{L}_{\text{sG}} \left[\varphi_1\right]
        + \mathcal{L}_{\text{sG}} \left[\varphi_2\right]
        + \frac{1}{2\pi K} \left(
            u \gamma_x \partial _x \varphi_2 \partial _x \varphi_1+ \frac{\gamma_\tau}{u} \partial _\tau \varphi_2 \partial _\tau \varphi_1
        \right),
\end{equation}
where $K, u$ are the Luttinger parameter and velocity of the sG Lagrangian $\mathcal{L}_{\text{sG}}$. $\gamma_{x,\tau}$ are quadratic coupling coefficients
(in terms of the parameters in Eq.~\eqref{eq: Phi lagrangian}, $\gamma_x = 2J_{+;l} / J_{+; 0}$ and $\gamma_\tau = 2J_{-;l} / J_{-; 0}$). We assume $|\gamma_{x,\tau}| < 2$ as required for stability.
Suppose $\gamma_{x} > 0$.

\emph{Semiclassical treatment.}
Let us look at the soliton--antisoliton saddle point and write $\varphi_1 = \varphi_c + \delta \varphi_1$, $\varphi_2 = -\varphi_c + \delta \varphi_2$, where $\varphi_c$ is the classical solution. We define $\delta \varphi_\pm = \left(\delta \varphi_1 \pm \delta \varphi_2\right) / \sqrt{2}$ and treat the system in the semiclassical approximation, keeping terms up to quadratic order in $\delta \varphi_{\pm}$. The  resulting Lagrangian reads

\begin{gather}
    \label{eq: quadratic split csg}
    \begin{aligned}
        \mathcal{L}_{\text{csG}} 
        \left[\varphi_c ; \delta\varphi_\pm\right]
        & = 2\mathcal{L}_{\text{eff-sG}}
        \left[\varphi_{\text{c}}\right]
        + \sum _{\alpha \in \{\pm\}} \left(
            \frac{1}{2\pi K}\left(u\left(1+\alpha\frac{\gamma_{x}}{2}\right)\left(\partial_{x}\delta\varphi_{\alpha}\right)^{2}+\frac{1}{u}\left(1+\alpha\frac{\gamma_{\tau}}{2}\right)\left(\partial_{\tau}\delta\varphi_{\alpha}\right)^{2}\right)
            + 4g\cos\left(\sqrt{8}\varphi_{c}\right)\delta\varphi_{\alpha}^{2}
        \right)
        \\ \mathcal{L}_{\text{eff-sG}}\left[\varphi_{\text{c}}\right] & = \frac{1}{2\pi K}\left(u\left(1-\frac{\gamma_{x}}{2}\right)\left(\partial_{x}\varphi_{c}\right)^{2}
        + \frac{1}{u}\left(1-\frac{\gamma_{\tau}}{2}\right)\left(\partial_{\tau}\varphi_{c}\right)^{2}\right)
        +g\left(1-\cos\left(\sqrt{8}\varphi_{c}\right)\right).
    \end{aligned}
\end{gather}

The classical soliton mass may be calculated directly from $2\mathcal{L}_{\text{eff-sG}}$. Quantum fluctuations contribute through the $\mathcal{O} \left(\delta \varphi_\pm ^2\right)$ terms of $\mathcal{L}_{\text{csG}}$, that may be integrated out.
For convenience we define
\begin{equation}
    K_\alpha = \frac{K}{\sqrt{\left(1 + \alpha \gamma_\tau / 2\right)\left(1 + \alpha \gamma_x / 2\right)}};
    \quad\quad\quad
    u_\alpha \equiv u\sqrt{\frac{1 + \alpha \gamma_x / 2}{1 + \alpha \gamma_\tau / 2}}.
\end{equation}
These differ from the main text Luttinger parameters, $K_{+/-} = 2K_{\text{2w;s/as;}l}$.
The fluctuation action now reads

\begin{gather}
    \begin{aligned}
        \mathcal{S}_\text{fluctuation}
        & = \sum _{\alpha \in \{\pm\}} \int _{\tau, x} \left(
            \frac{1}{2\pi K_\alpha}\left(
                \frac{1}{u_\alpha}\left(\partial_{\tau}\delta\varphi_{\alpha}\right)^{2}
                + 
                u_\alpha\left(\partial_{x}\delta\varphi_{\alpha}\right)^{2}
            \right)
            + 4g\cos\left(\sqrt{8}\varphi_{c}\right)\delta\varphi_{\alpha}^{2}
        \right)
        \\ & = \sum _{\alpha \in \{\pm\}} \int _{\tau, x} \delta\varphi_{\alpha} \left[
            - \frac{1}{2\pi K_\alpha}\left(
                \frac{1}{u_\alpha}\partial_{\tau}^2
                + u_\alpha\partial_{x}^2
            \right)
            + 4g\left(1 - 2 \text{sech} ^2 \left(\frac{x}{L_-}\right)\right)
        \right] \delta\varphi_{\alpha}
    \end{aligned},
\end{gather}
where we integrated by parts the first terms and substituted the soliton solution in the last. $L_- \equiv \sqrt{u_-/ 8\pi g K_-}$ is the soliton width of $\mathcal{L}_{\text{eff-sG}}$. We rescale, $x \equiv L_- \xi$ and, separately for each $\alpha$, $\tau \equiv \left(L_- / u_\alpha\right)\theta$, to integrate the quadratic action and obtain
\begin{equation}
    \Delta \mathcal{S}_\alpha = \frac{1}{2} \text{Tr} \ln \left[
            - \partial_{\theta}^2  
            - \partial_{\xi}^2
            + r_\alpha \left(1 - 2 \text{sech} ^2 \xi\right)
    \right] - \frac{1}{2} \text{Tr} \ln \left[
            - \partial_{\theta}^2  
            - \partial_{\xi}^2
            + r_\alpha
    \right];
    \quad\quad\quad
    r_\alpha \equiv \frac{1 - \gamma_x / 2}{1 + \alpha \gamma_x / 2},
\end{equation}
where $\Delta \mathcal{S}_\text{fluctuation} = \Delta \mathcal{S}_+ + \Delta\mathcal{S}_-$.
The operator $-\partial_\xi ^2 + r \left(1 - 2 \text{sech} ^2 \xi\right)$ is a P{\"o}schl--Teller operator~\cite{poeschlBemerkungenZurQuantenmechanik1933}. Its log determinant is given by~\cite{dunneFunctionalDeterminantsQuantum2008, dunne_integer}
\begin{equation}
    \label{eq: poschl-teller-log-det}
    \ln \left(\frac{\det \left[\mu^2 - \partial_\xi ^2 - \ell\left(\ell + 1\right)\text{sech} ^2 \xi\right]}{\det \left[\mu^2 - \partial_\xi ^2\right]}\right) = \ln \left(\frac{\Gamma\left(1 + \mu\right) \Gamma\left(\mu\right)}{\Gamma\left(1 + \mu + \ell\right) \Gamma\left(\mu - \ell\right)}\right)
\end{equation}
where we use, following the common convention, $2r \equiv \ell\left(\ell + 1\right)$, replace $-\partial_\theta ^2$ with its eigenvalue $\omega^2$ and define $\mu = \sqrt{r + \omega ^2}$.
The summation over all eigenvalues is UV log-divergent: asymptotically Eq.~\eqref{eq: poschl-teller-log-det} is $\left(\int d\xi V\left(\xi\right) \right) / 2\mu = -\ell\left(\ell + 1\right) / \mu$ where $V\left(\xi\right) = -\ell\left(\ell + 1\right) \text{sech} ^2 \xi$ is the potential. This first-order term is the tadpole,
which is removed by normal ordering~\cite{grahamUnambiguousOneloopQuantum1998}. We therefore add $\ell\left(\ell + 1\right) / \mu$ to the integrand. Introducing a finite time $T$ for the action ($\tau \in \left[0, T\right]$), we get

\begin{equation}
    \label{eq: Delta Salpha}
    \Delta \mathcal{S}_\alpha / T
        = \frac{u_\alpha}{L_-}
        \cdot \frac{1}{2}\int\frac{d\omega}{2\pi}
        \left(
            \ln \left(\frac{\Gamma\left(1 + \mu_\alpha\left(\omega\right)\right) \Gamma\left(\mu_\alpha\left(\omega\right)\right)}{\Gamma\left(1 + \mu_\alpha\left(\omega\right) + \ell_\alpha\right) \Gamma\left(\mu_\alpha\left(\omega\right) - \ell_\alpha\right)}\right)
            + \frac{\ell_\alpha\left(\ell_\alpha + 1\right)}{\mu_\alpha\left(\omega\right)}
        \right),
\end{equation}
where the last term fixes the UV divergence. 
Let $M_{-;\text{c}}$ be the classical soliton mass of $\mathcal{L}_{\text{eff-sG}}$, and notice that $u_\alpha / L_- = \pi M_{-;\text{c}} K_- \left(u_\alpha / u_-\right)$ (see Appendix~\ref{appx: sine Gordon}). For $\alpha = -$, $r_- = \ell_- = 1$ and the integral greatly simplifies. We find
\begin{equation}
    \Delta \mathcal{S}_- / T
    = M_{-;\text{c}} K_- \cdot
    \frac{\pi}{2}
    \underbrace{\int\frac{d\omega}{2\pi}
        \left(
            \ln \left(\frac{\sqrt{1 + \omega^2} - 1}{\sqrt{1 + \omega^2} + 1}\right)
            + \frac{2}{\sqrt{1 + \omega^2}}
        \right)
    }_{=-2/\pi} = - M_{-;\text{c}} K_-,
\end{equation}
precisely the DHN shift of the sG model~\cite{dashenParticleSpectrumModel1975, rajaraman1982solitons}. 
Fluctuations in the $\alpha = +$ channel will increase this shift, yielding the soliton--antisoliton mass $M_{\text{s-as}} = M_{-;\text{c}} \left(2 - K_-\left(1 + \delta_{\text{harmonic}}\right)\right)$,
where $\delta_{\text{harmonic}} = -\Delta \mathcal{S}_+ / \left(TM_{-;\text{c}} K_-\right)$.
For $\alpha=+$, $0 < r_+ < 1$, $\ell$ is non-integer and thus Eq.~\eqref{eq: Delta Salpha} has to be evaluated numerically. 

\emph{Pseudo-Goldstone mode.}
The formula for $\Delta \mathcal{S}_+$ requires a modification. Consider the decoupled limit ($\gamma_x = \gamma_\tau = 0$), which has two Goldstone zero modes relating to the independent translation of solitons in each sG theory. These modes are not well described by the harmonic treatment of the semiclassical approximation. Nonetheless, they are benign and do not contribute to the mass. Turning on the coupling, the symmetry is reduced, leaving us with a Goldstone zero mode for center-of-mass translation in $\mathcal{S}_-$ and a gapped, pseudo-Goldstone mode for relative translation in $\mathcal{S}_+$. 
The latter is not benign. At leading order in $\gamma_{x}$, it is $\delta_{\text{harmonic}} \approx 1 - \left(\pi/2\sqrt{3}\right)\sqrt{\gamma_x}$. For sufficiently small couplings, it outweighs the classical $\mathcal{O}\left(\gamma_{x}\right)$ binding energy, yielding $M_{\text{s-as}} > 2 M_{\text{s}}$ with $M_{\text{s}}$ the single soliton mass. This would imply there is no bound state, contrary to Ref.~\cite{simonBoundStateWeakly1976} and disqualify the semiclassical approximation.
As we will soon see, the leading $\sqrt{\gamma_x}$ term in $\delta_{\text{harmonic}}$ is a result of the pseudo-Goldstone mode. Thus, its inclusion in $\delta_{\text{harmonic}}$ is not only unjustified in the semiclassical approximation but also qualitatively erroneous. Therefore,
we calculate it without relying on the semiclassical approximation. First,
we remove it from $\delta_\text{harmonic}$, and then re-add it by introducing a collective coordinate.

\emph{Pseudo-Goldstone removal.}
For $0 < r < 1$ the P{\"o}schl--Teller operator has exactly one bound state at $\lambda_0 = \ell - r$~\cite{landauQM_sech2}. It becomes a zero mode both in the decoupled limit and for $\alpha=-$, where $r=\ell=1$. It thus suffices to remove 
\begin{equation}
    \Delta \mathcal{S}_{\alpha} ^{\text{remove}} \equiv \frac{1}{2} \text{Tr} \ln \left[-\partial_\theta^2 + \lambda_0\right] - \frac{1}{2}\text{Tr} \ln \left[-\partial_\theta ^2\right]
    = T\frac{u_\alpha}{L_-}\frac{1}{2} \int \frac{d\omega}{2\pi} \ln\left(\frac{\omega^2 + \lambda_0}{\omega^2}\right) = T\frac{u_\alpha}{L_-} \cdot \frac{1}{2}\sqrt{\lambda_0}.
\end{equation}
The removed vacuum value is in accordance with our convention for which 
$\Delta \mathcal{S}_{\alpha} ^{\text{remove}} = 0$ for the zero mode ($\lambda_0 = 0$).
Using $\omega^2 + \lambda_0 = \mu^2 - \ell^2$, we write
\begin{equation}
    \delta_{\text{no-pG}} = - \frac{u_+}{u_-} \cdot \frac{\pi}{2} \int\frac{d\omega}{2\pi}
        \left(
            \ln \left(\frac{\omega^2\Gamma\left(1 + \mu_+\left(\omega\right)\right) \Gamma\left(\mu_+\left(\omega\right)\right)}{\left(\mu_+\left(\omega\right) + \ell_+\right)\Gamma\left(1 + \mu_+\left(\omega\right) + \ell_+\right) \Gamma\left(1 + \mu_+\left(\omega\right) - \ell_+\right)}\right)
            + \frac{\ell_+\left(\ell_+ + 1\right)}{\mu_+\left(\omega\right)}
        \right).
\end{equation}
At weak coupling, $\delta_{\text{no-pG}} \approx 1 + \gamma_x / 6 - \gamma_\tau / 2$. The non-analytic square root in $\delta_{\text{harmonic}}$ originates from $\Gamma\left(\mu - \ell\right)$, which cannot be expanded around $\mu - \ell = 0$ at $\omega=0$ and weak coupling. In the $\delta_{\text{no-pG}}$ integrand, analytic expansion can be done safely.

\emph{Collective coordinate.}
Lastly, we need to take into account the pseudo-Goldstone mode beyond the semiclassical approximation, which may be done by means of a collective coordinate. 
We have already computed the zero point motion of the non-soft modes at zero separation. Below, we derive an effective Hamiltonian for the soliton--antisoliton separation R, assuming it is slow enough to decouple from fluctuations in the harmonic channels.
We substitute in $\varphi_1 = \varphi_{c} \left(x - X_1 \left(t\right)\right)$, $\varphi_2 = -\varphi_{c} \left(x - X_2 \left(t\right)\right)$ into the real-time version of the Lagrangian density in Eq.~\eqref{eq: two coupled sG Lagrangian}, and integrate over space.
Here, $\varphi_c$ is a soliton of $\mathcal{L}_{\text{sG}}$ and $L \equiv \sqrt{u / 8\pi gK}$ its width. We get a Lagrangian for the soliton positions $X_1, X_2$,

\begin{equation}
    \mathfrak{L}_\text{cc} \left[X_1, X_2\right]
    = \frac{1}{2}M_{\text{0c}} \left(
        \frac{1}{u^2} \left(\dot{X}_1 ^2 + \dot{X}_2 ^2\right)
        -\frac{\gamma_\tau }{u^2} \frac{\left(X_1 - X_2\right) / L}{\sinh\left(\left(X_1 - X_2\right) / L\right)} \dot{X_1} \dot{X_2}
        +\gamma_x \frac{\left(X_1 - X_2\right) / L}{\sinh\left(\left(X_1 - X_2\right) / L\right)}
    \right) - 2M_{\text{0c}},
\end{equation}
with $M_{\text{0c}}$ the classical soliton mass of $\mathcal{L}_{\text{sG}}$ ($M_{-\text{;c}} = M_\text{0c} \sqrt{1 - \gamma_x / 2}$).
We may define the center-of-mass coordinate $X \equiv \left(X_1 + X_2\right)/2$ and the relative coordinate $R = X_1 - X_2$. It is easy to see that
$\mathfrak{L}_\text{cc}$ does not depend on $X$ explicitly, which corresponds to a center-of-mass translation symmetry.
We consider the dipole at rest, set $X = \text{const}$, and get an effective Lagrangian for $R$,
\begin{equation}
    \mathfrak{L}_\text{cc} \left[R\right] = \frac{1}{2} M_{\text{0c}} \left(
        \frac{1}{2u^2}\left(1 + \frac{\gamma_\tau }{2} \frac{R / L}{\sinh\left(R / L\right)}\right) \dot{R} ^2
        +\gamma_x \frac{R / L}{\sinh\left(R / L\right)}
    \right) - 2M_{\text{0c}}.
\end{equation}
The corresponding Hamiltonian reads
\begin{equation}
    \mathfrak{H}_{\text{cc}} = P \frac{1}{2\mathfrak{m}\left(R\right)} P + U\left(R\right);
    \quad\quad
    U\left(R\right) \equiv 2M_{\text{0c}}\left(1 - \frac{\gamma_x}{4}  \frac{R / L}{\sinh\left(R / L\right)}\right);
    \quad\quad
    \mathfrak{m}\left(R\right) \equiv \frac{M_{\text{0c}}}{2u^2} \left(1 + \frac{\gamma_\tau}{2} \frac{R/L}{\sinh\left(R/L\right)}\right),
\end{equation}
with $P$ the conjugate momentum to $R$. We note that the classical coupling implied by $U\left(R\right)$ agrees with the classical one from Eq.~\eqref{eq: two-wire excitation energy} to first order in $\gamma_x$. We can now diagonalize $\mathfrak{H}_{\text{cc}}$ and compute its zero-point-motion energy $E_{\text{cc;0}}$ (above the classical minimum), shifting the mass and yielding
\begin{equation}
    \delta = \delta_{\text{no-pG}} - \frac{E_{\text{cc;0}}}{K_- M_{-\text{;c}}}.
\end{equation}
It is this formula we use to calculate the shift of the symmetric mode fluctuations ($\delta_l$ of Eq.~\eqref{eq: two-wire excitation energy}, which is used in Figs.~\ref{fig: branch sketch} and~\ref{fig: FCI magnetorotons}).
We note that $E_{\text{cc;0}}$ is analytic in $\gamma_x$~\cite{simonBoundStateWeakly1976} and therefore cannot re-introduce the $\mathcal{O}\left(\sqrt{\gamma_x}\right)$ term.

\section{Soliton dressing}
\label{appx: dressing}

When deriving the effective sG models $\mathcal{L}_{\text{1w}}$ and $\mathcal{L}_{\text{2w;as;}l}$ in Section~\ref{sec: neutral exc}, we neglected subleading quadratic  contributions in the inter-wire coupling ($\sim \mathcal{O}\left(\left(J_{\pm;l} / J_{\pm;0}\right)^2\right)$) that arise due to the presence of additional ``idle'' wires that do not participate in the sG model but are nonetheless coupled to it. In this appendix, we derive these corrections and show that they are indeed quadratic and hence subleading.

For decoupled wires, excitations of $\mathcal{L}_{\Phi}$ (Eq.~\eqref{eq: Phi lagrangian}) are composed of solitons in each wire. For non-zero but weak coupling, we expect the solitons to be dressed by their interactions with the other wires. Namely, when a soliton occurs in a wire, other wires will experience a ``bump'' ($\Phi \neq 0$), which may affect the properties of the soliton. Bump sizes will be proportional to the coupling, and hence, in leading order, we may assume bumps do not interact with one another. This allows us to restrict ourselves to the much simpler two-wire problem (two coupled sG theories), then sum over contributions from different bumps. 

Let $\varphi_1$ be the field on a wire hosting a soliton, and $\varphi_2$ be the field on a wire hosting a bump. The ``soliton + bump'' Lagrangian, $\mathcal{L}_{\text{s+b}}$, is then given by Eq.~\eqref{eq: two coupled sG Lagrangian} where we approximate $1 - \cos\left(\sqrt{8} \varphi_2\right) \approx 4 \varphi_2^2$,
\begin{gather}
    \begin{aligned}
        \mathcal{L}_{\text{s+b}} \left[\varphi_1, \varphi_2\right] = \mathcal{L}_{\text{sG}} \left[\varphi_1\right]
        + \frac{1}{2\pi K} \left(
            u \left(\partial _x \varphi_2\right) ^2 + \frac{1}{u} \left(\partial _\tau \varphi_2\right) ^2 
        \right) + 4g\varphi_2 ^2
        + \frac{1}{2\pi K} \left(
            u \gamma_x \partial _x \varphi_2 \partial _x \varphi_1+ \frac{\gamma_\tau}{u} \partial _\tau \varphi_2 \partial _\tau \varphi_1
        \right).
    \end{aligned}
\end{gather}
The Lagrangian is quadratic in $\varphi_2$, and we may complete the square to integrate it out, leaving the residual action
\begin{equation}
    \Delta \mathcal{S} = -\frac{1}{8\pi K} \frac{1}{\beta\Omega}\sum_{\boldsymbol{q}}\frac{\left(\gamma_x u q^2 + \gamma_\tau u^{-1}
    \omega_n ^2\right)^2}{uq^2 + u^{-1}\omega_n ^2 + 8\pi Kg} \varphi_{1;\boldsymbol{q}} \varphi_{1;-\boldsymbol{q}} 
\end{equation}
where $\boldsymbol{q} = \left(q, i\omega_n\right)$ and $\omega_n$ are the Matsubara frequencies. $\beta$ is the inverse temperature and $\Omega$ the system size. The correction is evidently $\mathcal{O} \left(\gamma^2\right)$, subleading to the $\mathcal{O}\left(\gamma\right)$ obtained in Eqs.~\eqref{eq: two-wire excitation energy},~\eqref{eq: quadratic split csg}, so in leading order we may neglect the effect of dressing by other wires.

\section{Fractional Chern insulator model}
\label{appx: FCI}
In this appendix, we provide some technical details for the results discussed in Section~\ref{sec: stability}.
The model considered in Section~\ref{sec: stability} was presented in
Ref.~\cite{shavitQuantumGeometryStabilization2024}, and takes the form of a two-wire bosonized Hamiltonian describing the competition of an FCI, an anti-FCI, and two CDW orders,
\begin{gather}
    \label{eq: two-wire hamiltonian}
    \begin{aligned}
        \mathcal{H}_{\text{2w-FCI}} 
        = & \frac{u}{2\pi} \sum _{\alpha=\rho,\sigma} \left(
            K_\alpha \left(\partial_x \theta_\alpha\right)^2
            + \frac{1}{K_\alpha} \left(\partial_x \phi_\alpha\right)^2
        \right)
        + \frac{\tilde{V} u}{2\pi}\left(\sqrt{\frac{K_{\sigma}}{K_{\rho}}}\partial_{x}\phi_{\rho}\partial_{x}\theta_{\sigma}-\sqrt{\frac{K_{\rho}}{K_{\sigma}}}\partial_{x}\phi_{\sigma}\partial_{x}\theta_{\rho}\right)
        \\ & + \frac{g_\text{FCI}}{\left(2\pi\right)^2}\mathcal{O}_{\text{FCI}} + \frac{g_\text{aFCI}}{\left(2\pi\right)^2}\mathcal{O}_{\text{aFCI}} 
        + \frac{g_\text{dCDW}}{\left(2\pi\right)^2}\mathcal{O}_{\text{dCDW}} + \frac{g_\text{gCDW}}{\left(2\pi\right)^2}\mathcal{O}_{\text{gCDW}}.
    \end{aligned}
\end{gather}
Here, $\rho$ and $\sigma$ denote the symmetric and antisymmetric components of the two wires, respectively: $\phi_{\rho,\sigma} = \left(\phi_1 \pm \phi_2\right) / \sqrt{2}$ and similarly for $\theta$.
The operators $\mathcal{O}_{\text{FCI}}, \mathcal{O}_{\text{aFCI}}, \mathcal{O}_{\text{dCDW}}$ and $\mathcal{O}_{\text{gCDW}}$ are given by
\begin{gather}
    \label{eq: FCI operator def}
    \begin{aligned}
        \mathcal{O}_{\text{FCI}} & = \cos\left(\sqrt{2}\theta_{\sigma}+m\sqrt{2}\phi_{\rho}\right)
        \\ \mathcal{O}_{\text{aFCI}} & = \cos\left(\sqrt{2}\theta_{\sigma}-m\sqrt{2}\phi_{\rho}\right)
        \\ \mathcal{O}_{\text{dCDW}} & = \cos\left(m\sqrt{2}\phi_{\rho}\right)\cos\left(m\sqrt{2}\phi_{\sigma}\right)
        \\ \mathcal{O}_{\text{gCDW}} & = \cos\left(\sqrt{8}\phi_{\sigma}\right).
    \end{aligned}
\end{gather}
$\mathcal{O}_{\text{FCI}}$ and $\mathcal{O}_{\text{aFCI}}$ stabilize the FCI and the anti-FCI, respectively. The operator $\mathcal{O}_{\text{FCI}}$ is simply $\mathcal{O}_{\text{Laughlin}}$ (Eq.~\eqref{eq: LaughlinOpr}) on the two wires, and $\mathcal{O}_{\text{aFCI}}$ takes the same form with a magnetic field in the opposite direction ($m$ replaced by $-m$). $\mathcal{O}_{\text{dCDW}}$ and $\mathcal{O}_{\text{gCDW}}$ stabilize charge-ordered states.
$\tilde V$ is a chiral interaction generated by the flow of $\mathcal{O}_{\text{FCI,aFCI}}$.

The second-order perturbative RG equations are derived using the operator product expansion (OPE)~\cite{cardyScalingRenormalizationStatistical1996}, as exemplified in the supplemental material of Ref.~\cite{shavitQuantumGeometryStabilization2024}. They may be derived following Appendices C and E of Ref.~\cite{giamarchiQuantumPhysicsOne2003} as well as the identification $\partial_\tau \phi = -ivK\partial_{x}\theta$, which follows immediately from Hamilton's equations for the Tomonaga-Luttinger Hamiltonian.
We define the dimensionless coupling constants $y_\alpha \equiv g_\alpha / 2\pi u$. The resulting flow equations read

\begin{gather}
    \label{eq: RG flow}
    \begin{aligned}
        \frac{d}{d\ell}y_{\text{FCI}}&=\left(2-d_{\text{FCI}}+\frac{m}{2}\sqrt{\frac{K_{\rho}}{K_{\sigma}}}\tilde{V}\right)y_{\text{FCI}}\\\frac{d}{d\ell}y_{\text{aFCI}}&=\left(2-d_{\text{FCI}}-\frac{m}{2}\sqrt{\frac{K_{\rho}}{K_{\sigma}}}\tilde{V}\right)y_{\text{aFCI}}\\\frac{d}{d\ell}y_{\text{dCDW}}&=\left(2-d_{\text{dCDW}}\right)y_{\text{dCDW}}\\\frac{d}{d\ell}y_{\text{gCDW}}&=\left(2-2K_{\sigma}\right)y_{\text{gCDW}}\\\frac{d}{d\ell}K_{\rho}^{-1}&=\frac{m^{2}}{8}\left(y_{\text{FCI}}^{2}+y_{\text{aFCI}}^{2}+\frac{1}{2}y_{\text{dCDW}}^{2}\right)\\\frac{d}{d\ell}K_{\sigma}&=\frac{1}{8}\left(y_{\text{FCI}}^{2}+y_{\text{aFCI}}^{2}\right)-K_{\sigma}^{2}\left(\frac{m^{2}}{16}y_{\text{dCDW}}^{2}+\frac{1}{2}y_{\text{gCDW}}^{2}\right)\\\frac{d}{d\ell}\tilde{V}&=\frac{m}{4}\sqrt{\frac{K_{\rho}}{K_{\sigma}}}\left(y_{\text{FCI}}^{2}-y_{\text{aFCI}}^{2}\right)
    \end{aligned}
\end{gather}
with the dimensions
\begin{gather}
    \begin{aligned}
        d_{\text{FCI}} = \frac{1}{2}m^{2}K_{\rho}+\frac{1}{2}K_{\sigma}^{-1}
        ;\quad\quad d_{\text{dCDW}} = \frac{1}{2}m^{2}K_{\rho}+\frac{1}{2}m^{2}K_{\sigma}.
    \end{aligned}
\end{gather}

$\mathcal{O}_{\text{FCI}}$ generates $\tilde{V}$, which in turn lowers the dimension of $\mathcal{O}_{\text{FCI}}$, further stabilizing the fractional state. $\mathcal{O}_{\text{aFCI}}$ has the opposite effect, hindering the formation of an FCI and possibly destabilizing it. We may write
\begin{gather}
    \begin{aligned}
        \sqrt{\frac{K_{\sigma}}{K_{\rho}}}
        \partial_x\phi_{\rho}\partial_x\theta_{\sigma}
        -\sqrt{\frac{K_{\rho}}{K_{\sigma}}}\partial_x\phi_{\sigma}\partial_x\theta_{\rho} = &\frac{1}{2}\left(\sqrt{\frac{K_{\rho}}{K_{\sigma}}}-\sqrt{\frac{K_{\sigma}}{K_{\rho}}}\right)\left(\partial_x\phi_{2}\partial_x\theta_{2}-\partial_x\phi_{1}\partial_x\theta_{1}\right)
        \\ & + \frac{1}{2}\left(\sqrt{\frac{K_{\rho}}{K_{\sigma}}}+\sqrt{\frac{K_{\sigma}}{K_{\rho}}}\right)\left(
        \partial_x\theta_{1}\partial_x\phi_{2}-\partial_x\phi_{1}\partial_x\theta_{2}\right)
    \end{aligned}
\end{gather}
In the multiwire system, the first line of the right-hand side cancels when summing over all wires. The second line is simply a $V_{\theta\phi;1}$ term in Eq.~\eqref{eq: bare Hamiltonian} ($\tilde{V} \propto V_{\theta\phi}$). Therefore, $V_{\theta\phi}$ terms in the IR theory, generated during the RG flow, constitute a good proxy for quantum geometry in the UV lattice model.

We note that the operator $\tilde{V}$ and the RG equations involving it differ from the ones presented in Ref.~\cite{shavitQuantumGeometryStabilization2024} (Eqs.~(9),~(S46)), primarily in $\sqrt{K_\rho / K_\sigma}$ factors. This is due to a minor error, likely a missing factor in the definition of the conjugate momenta ($\partial_\tau \phi = v\partial_x \theta$ instead of $\partial_\tau \phi = -ivK\partial_{x}\theta$). Upon rescaling $\tilde{V} \to \tilde{V} \sqrt{K_\rho / K_\sigma}$, the flow equations agree to quadratic order in the coupling and the results are unaffected.
\end{document}